# Synthesis and Magnetic Properties of Ferroelectric GdCrO$_3$ nanoparticles

Y. Sundarayya[1,2] and S. Srinath[1,*]
[1] School of Physics, University of Hyderabad, Hyderabad - 500046, A.P, India.
[2] Department of Physics, School of Sciences, Nagaland University, Lumami – 798627, Zunheboto, Nagaland, India.
[*] Corresponding author: sssp@uohyd.ernet.in

**Abstract:** Homogeneous single phase GdCrO$_3$ nanoparticles are synthesized by a modified-hydrothermal synthesis. The sample shows a compensation temperature at 128 K, below which the DC magnetization becomes negative and positive at low temperatures due to the competition between the two sublattice magnetization. At Néel temperature (168K), the line width and the intensity show an abrupt transition, revealed from electron paramagnetic resonance spectroscopy.

## 1. INTRODUCTION

Rare-earth orthochromites, RCrO$_3$ (R = rare earth/ Yttrium) is one class of oxides that got the attention of researchers due to their electronic correlations. When R is non-magnetic, the magnetic structure is solely governed by Cr–O–Cr superexchange interaction which results in the Cr sublattices order antiferromagnetically but with a canting of spins at Cr sites or weak ferromagnetism below Néel temperature ($T_{N1}$ ~ 200 K), attributed to antisymmetric Dzyaloshinsky-Moriya (DM) exchange interaction between neighbouring spins. When R-ions are magnetic, there are three kinds of interactions, namely Cr$^{3+}$-Cr$^{3+}$, R$^{3+}$-Cr$^{3+}$ and R$^{3+}$-R$^{3+}$ which result in a complex magnetic behaviour [1]. The R$^{3+}$-Cr$^{3+}$ interactions result in several interesting properties such as spin-reorientation and magnetization reversal below a compensation temperature [2, 3]. Recently, field-induced ferroelectricity at Néel temperature has been reported in weakly ferromagnetic RCrO$_3$, indicating a strong magnetoelectric effect [4]. However, it is expected that the dynamics of this mechanism greatly depends on size. We report here, the synthesis and magnetic properties of GdCrO$_3$, through DC magnetization studies and Electron spin resonance spectroscopy (ESR). ESR is a powerful tool to study the dynamics of spin states across the transition in a magnetic system, canted antoferromagnets in particular.

## 2. SYNTHESIS

Homogeneous single phase GdCrO$_3$ nanoparticles have been synthesized using a sol-gel followed by hydrothermal method [5]. In the present case, the stoichiometric amounts of chromium nitrate (Cr(NO$_3$)$_3$.9H$_2$O (Alfa-aesar, 99.9%), gadolinium nitrate hydrate Gd(NO$_3$)$_3$.3H$_2$O, (Alfa-aesar, 99.9%), and citric acid (Merck, 99.5%) with 1:1:1 molar ratio were dissolved in the deionized water so that the metal ions can be completely complexed to the citrate ions. This was followed by the drop-wise addition of ammonia solution (25 wt %) to raise the pH value of the solution to reach 9-10 resulting in a sol formation. The sol was transferred to a 200 ml capacity autoclave with the Teflon liner and was subjected to hydrothermal treatment at 473 K for 24 h. The precipitate was, in turn, filtered, washed with deionized water and dried at 423 K for 24 h. A pellet of the as-synthesized powder was annealed at 973 K in air for 12 h to obtain GdCrO$_3$ nanoparticles.

## 3. CHARACTERIZATION

The as-synthesized sample was characterized by X-ray powder diffraction (XRD) using Bruker D8 Advance diffractometer with Cu-K$\alpha$ radiation ($\lambda$ = 1.54056 Å). The morphology of the samples was studied by means of field-emission scanning electron microscopy (CARL ZEISS ULTRA-55 FESEM). Magnetic measurements were carried out with a vibrating sample magnetometer in a Physical Property Measurement System (PPMS). Temperature dependent electron paramagnetic resonance (EPR) spectra were recorded in the X-band region with frequency 9.155 GHz on JEOL-FE3X ESR spectrometer in the range 123 – 373 K, using a liquid nitrogen (LN2) cryostat.

## 4. RESULTS AND DISCUSSION

The X-ray diffractogram, of the as synthesized sample with Rietveld profile matching is shown in figure 1, confirms that the sample crystallizes into orthorhombic structure of GdCrO$_3$ with space group *Pnma*. The obtained lattice parameters are $a$ = 5.5447 (1) Å, $b$ = 7.6109 (1) Å and $c$ = 5.3184 (1) Å. The average crystallite size estimated from Scherer's formula [6] is found to be 35 (1) nm. The inset of Fig.1 shows the morphology of the GdCrO$_3$ and depicts the average particle size as 50 nm.

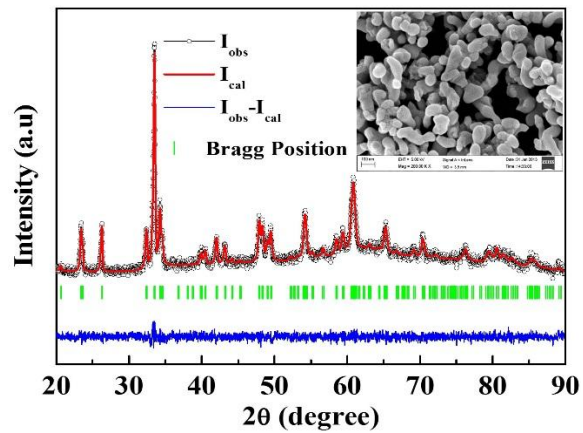

Figure 1. Powder X-ray diffractogram of the GdCrO$_3$ nanoparticles. Morphology by FESEM is shown in the inset.

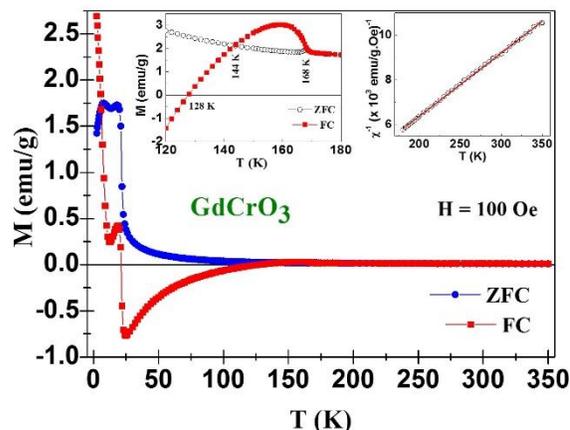

Figure 2. Zero field-cooled and field-cooled magnetization of GdCrO$_3$ nanoparticles with temperature. The inset shows the spin-reorientation, compensation temperature observed. The inset on the right shows the C-W analysis of 1/$\chi$ vs. T for T > $T_N$.

Figure 2 shows the zero field-cooled (ZFC) and field-cooled (FC) magnetization measurements on GdCrO$_3$ nanoparticles in the presence of an external magnetic field of 100 Oe. It shows the GdCrO$_3$ undergoes a paramagnetic to antiferromagnetic transition at ~ 168 K, the Néel temperature ($T_N$) of the Cr$^{3+}$ sublattice. The observed $T_N$ of the GdCrO$_3$ nanoparticles is close to the reported $T_N$ (167 K) of the bulk sample [4], confirms the Néel temperature does not get affected by the size of the GdCrO$_3$ nanoparticles in discussion. However, it is observed that the magnetization of the GdCrO$_3$ nanoparticles is not analogues to the bulk GdCrO$_3$. Both Zero field-cooled and field-cooled magnetization of the GdCrO$_3$ nanoparticles shows anomalous behaviour. Below the T$_N$, the FC increases initially and undergoes a spin-reorientation (SR) with a compensation temperature, T* (~ 128 K), as shown in the inset of Fig.2. Below 128 K, the sample undergoes a negative magnetization and exhibits further anomalies at 25, 19 and 13 K before it shoots up to attain saturation value 2.69 emu/gm at 2.5 K. This is due to the competition between the sub-lattice magnetization of Chromium and Gadolinium sub-lattices, and is expected in oxide materials. The inset in Figure 2 shows that the inverse susceptibility of GdCrO$_3$ in the paramagnetic region follows a typical Curie-Weiss (CW) behaviour. By fitting the CW law to $\chi^{-1}(T)$, we obtain the characteristic temperature θ = -25.9 (0.2) K. This confirms the presence of antiferromagnetic spin correlations in the compound. The effective magnetic moment (μ$_{eff}$), deduced from the Curie constant, for GdCrO$_3$ is 8.966μ$_B$. These values of θ and μ$_{eff}$ are in agreement with the reported values [7]. Figure 3 shows the magnetization curve, M(H) of the GdCrO$_3$ nanoparticles with external magnetic field at 2.5 K and inset shows the magnetization at 300 K. It may be concluded that the presence of pinching shows the presence of antiferromagnetic ordering in the sample; the non-zero hysteresis confirms the presence of canting associated with Cr$^{3+}$ sub-lattice. The observed coercive field (H$_c$) at 2.5 K is found to be 240 Oe.

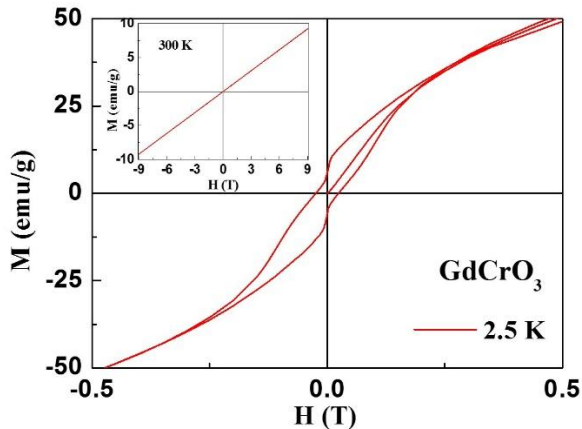

Figure 3. Magnetization curve, M(H) of the GdCrO$_3$ nanoparticles with external magnetic field at 2.5 K and 300 K (shown in inset).

To examine the paramagnetic to antiferromagnetic transition, Temperature dependent EPR spectra were recorded on GdCrO$_3$ nanoparticles in the X-band region in the range 123 – 373 K. Figure 4 shows the electron paramagnetic resonance spectrum of GdCrO$_3$ nanoparticles at 373 K, with rate of power *dP/dH* absorbed on the Y-axis with respect to the external magnetic field H on the X-axis. The inset shows the power absorbed with respect to external field, obtained by integrating the *dP/dH*. The observed data is shown by open circles. The observed values for the resonance field H$_{res}$ and the line-width ΔH$_{pp}$ (also, peak to peak width) at 373 K are 328 and 132 mT respectively. The value of the g-factor at 373 K is 1.992 and confirms the paramagnetic nature of GdCrO$_3$ nanoparticles. To ascertain whether or not the observed values resembles to true, the power curve was analysed by fitting it to one or more Gaussian and/or Lorentzian line shape, to obtain the resonance field H$_{res}$ and line-width ΔH$_{pp}$.

It is observed that the power curve at 373 K could be fitted better by a pair of Gaussian curves, as shown in the inset of Fig.4 by a relatively broad green and narrow blue curves. These two subspectra may correspond to paramagnetic Gd$^{3+}$ and Cr$^{3+}$ species in the GdCrO$_3$ nanoparticles. The red curve shows the resultant of the fit and nearly overlaps with observed pattern, thus confirms the validity of the fit. The values of H$_{res}$ so obtained are 357, 324 mT and of ΔH$_{pp}$ are 299 and 140 mT respectively. The values of g-factor for the two independent Gaussians obtained are 2.018 and 1.830 respectively. It is evident from Gaussian line shape of the sub spectra that the dipoles of Cr$^{3+}$ and Gd$^{3+}$ experience slightly different effective magnetic fields and only a fraction of the spins is in resonance with sweeping external magnetic field [8].

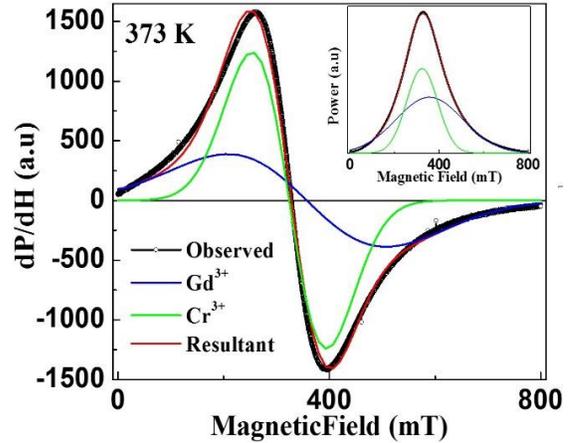

Figure 4. The EPR spectrum of GdCrO$_3$ nanoparticles recorded at 373 K. The deconvoluted components of paramagnetic Gd$^{3+}$ and Cr$^{3+}$ are shown by blue and green curves. The inset shows the power absorption with respect to the field.

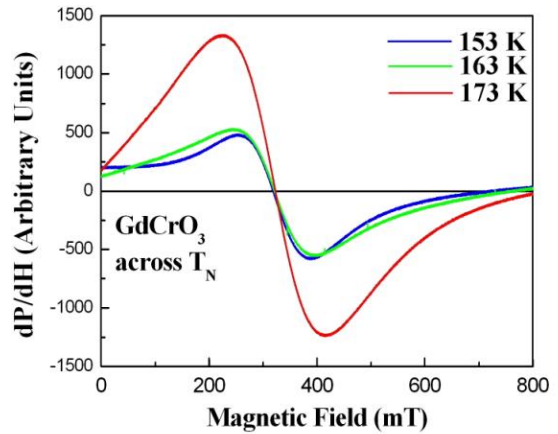

Figure 5. The EPR spectra of GdCrO$_3$ nanoparticles across Néel temperature.

Figure 5 shows the EPR spectra at 153, 163, and 173 K across the T$_N$ (168K). It is observed that the peak-to-peak amplitude (or intensity), I$_{pp}$ exhibits a decrease with decrease in temperature and undergoes an abrupt fall across the magnetic transition, as revealed from Fig. 5. The values of H$_{res}$ vary from 328 mT at 373 K to 319 mT at 123 K. The g-factor changes from 1.994 at 373 K to 2.050 at 123 K and corresponds to the paramagnetic nature of the sample.

However to understand the magnetic transition, the spectra in the temperature range 123 – 373 were analysed as mentioned above by a pair of Gaussians. The values of H$_{res}$, ΔH$_{pp}$ and g-factor so obtained are plotted as shown in Figure 6. It is evident from the figure that the values of g-factor for one set of values varies from 2.018 at 373 K to 2.012 at 123 K, resembles the paramagnetic nature of the sub-lattice. The second set values vary from 1.830 at 373 K to 2.242 at 123 K with an abrupt increase across the transition. It may conclude that the former set corresponds to the paramagnetic $Gd^{3+}$ and the latter to the $Cr^{3+}$ sub-lattice. It is evident from the Fig.6 that the $H_{res}$ values of paramagnetic $Gd^{3+}$ experience a smooth change, while the $H_{res}$ values of $Cr^{3+}$ spins undergo a sudden decrease across the transition with decrease in temperature. The $\Delta H_{pp}$ of both $Gd^{3+}$ and $Cr^{3+}$ increases smoothly from 373 K to 173 K. A sudden decrease of $\Delta H_{pp}$ values is observed across the transition.

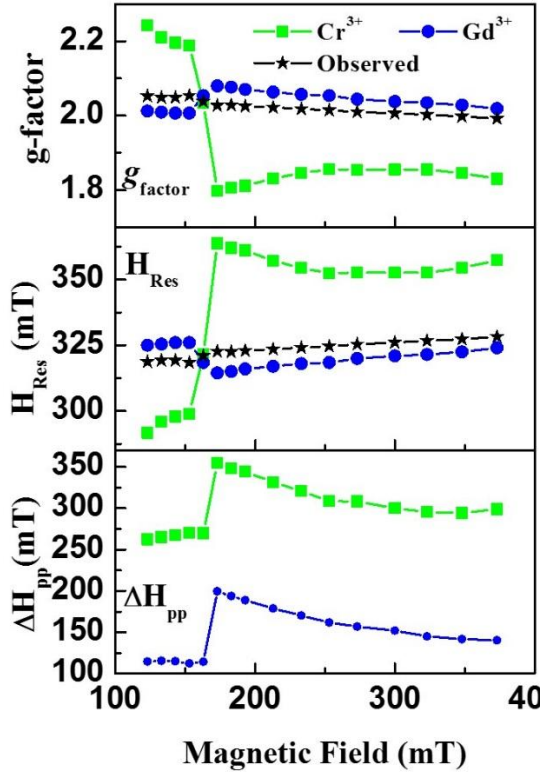

Figure 6. Variation of $\Delta H_{pp}$, $H_{res}$ and g-factor of $GdCrO_3$ nanoparticles with temperature.

## 4. CONCLUSIONS

$GdCrO_3$ nanoparticles have been synthesized by a modified so-gel followed by hydrothermal synthesis. These nanoparticles exhibit spin-orientation and negative magnetization below $T_N$ unlike the bulk $GdCrO_3$. Detailed analysis of the temperature dependent EPR spectra recorded in the range 123 – 373 K shows the $Gd^{3+}$ spins remains in the paramagnetic state while the $Cr^{3+}$ spins undergo a paramagnetic to antiferromagnetic transition at 168 K, corroborated by the variation in the values of $H_{res}$ and g-factors.

**ACKNOWLEDGEMENTS**
One of the authors YS thanks the DST-PURSE program for the postdoctoral fellowship. The authors thank Prof. S.N. Kaul for discussions and acknowledge the centre for nanotechnology (CFN), University of Hyderabad for magnetic measurements.